\documentclass[12pt,authoryear]{elsarticle}
 \usepackage{amssymb}
 \usepackage{natbib}
\usepackage{amsfonts,amsmath,epsfig,psfrag}
 \usepackage{amsmath}
 \usepackage{tikz}
 \usepackage{float}
\usepackage{pgflibraryshapes}
 \usepackage{graphicx,psfrag}
\usepackage[normalem]{ulem}
 \makeatletter \oddsidemargin 0.0cm \evensidemargin 0.0cm
\newcommand{\be}{\begin{equation}}
\newcommand{\en}{\end{equation}}
\newcommand{\la}{\label}
\newcommand{\ep}{{\varepsilon}}
\newcommand{\paa}{\partial}
\def\rr#1{(\ref{#1})}

\newcommand{\s}[1]{{\Large\textsf{\textbf{#1}}}}

\begin{document}
\begin{frontmatter}
\title{\s{Localised bulging of an inflated rubber tube with fixed ends}}
\author[mymainaddress]{Zhiming Guo }
\author[secondaryaddress]{Shibin Wang }
\author[thirdaddress]{Yibin Fu\corref{mycorrespondingauthor}}
\cortext[mycorrespondingauthor]{Corresponding author: y.fu@keele.ac.uk}
\address[mymainaddress]{College of Electromechanical Engineering, Changsha University, China}
\address[secondaryaddress]{Department of Mechanics, Tianjin University, China}
\address[thirdaddress]{School of Computing and Mathematics, Keele University, Staffs ST5 5BG, UK}

\begin{abstract}
When a rubber tube with free ends is inflated under volume control, the pressure will first reach a maximum and then decrease monotonically to approach a constant asymptote. The pressure maximum corresponds to the initiation of a localised bulge and is predicted by a bifurcation condition, whereas the asymptote is the Maxwell pressure corresponding to a \lq\lq two-phase" propagation state. In contrast, when the tube is first pre-stretched and then has its ends fixed during subsequent inflation, the pressure versus bulge amplitude has both a maximum and a minimum, and the behaviour on the right ascending branch has previously not been fully understood. We show that for all values of pre-stretch and tube length, the ascending branches all converge to a single curve that is only dependent on the ratio of tube thickness to the outer radius. This curve represents the Maxwell state to be approached in each case (if Euler buckling or axisymmetric wrinkling does not occur first), but this state is pressure-dependent in contrast with the free-ends case.  We also demonstrate experimentally that localized bulging cannot occur when the pre-stretch is sufficiently large and investigate what strain-energy functions can predict this observed phenomenon.
\end{abstract}

\begin{keyword}
Bifurcation\sep Stability\sep Rubber tubes\sep Localisation\sep Inflation
\end{keyword}
\end{frontmatter}

\section{Introduction}
A rubber tube or balloon may be inflated in two different ways. In the first scenario, the ends are closed but free to move in the axial direction. A dead weight may also be attached to one end in order to exert a constant axial force during inflation. In the other scenario which is the focus of the current study, the tube is first subjected to a pre-stretch and then the two ends are fixed during subsequent inflation.

Localised bulging in an inflated rubber tube is one of the oldest elasticity problems that is still being studied today due to its theoretical importance and applications. On the one hand, the primary deformation is an archetypal problem in nonlinear elasticity: it can be solved exactly and the resulting solution can be used to validate constitutive assumptions \citep{rivlin1949, gent-rivlin1952}. On the other hand, localised bulging of a balloon can routinely be observed in everyday life and mathematically can be viewed as one of the simplest sub-critical bifurcation problems. Also, the localisation process exhibits three distinct phases:  initiation, growth and propagation, which are also shared by a large variety of other localization problems in continuum mechanics (e.g phase transformations). The earliest study on localised bulging was by \citet{ma1891} who attempted to use linear elasticity to describe the condition for its appearance. Subsequent major studies include the experimental investigations by
 \citet{kc1990, kc1991},   \citet{pgl2006}, and \citet{gpl2008}, the numerical simulations by  \citet{sm1996}, and the theoretical studies by  \citet{yi1977} and \citet{ch1984} who recognised that the propagation stage for the free-ends case corresponds to a \lq\lq two-phase" deformation governed by Maxwell's equal area rule. The problem has received renewed interest in the recent decade due to its relevance to a variety of applications, as witnessed by a series of recent studies on the continuum-mechanical modelling of aneurysm initiation in human arteries \citep{frz2012, agrm2013,bh2013a, bh2013b, arm2014, rfz2015}, on localized bulging under the additional effects of swelling \citep{dm2015}, viscoelasticity/chemorheology \citep{wi2015a, wi2017}, and electric actuation \citep{laly2015}, on the development of 1D gradient models and analytical solutions \citep{LA18, LA20b, giudici2020}, and on mechanisms of rupture \citep{ hhp2021}. Inflated soft tubes are also increasingly being used in soft robotics \citep{usevitch2020, jin2021}.

An inflated rubber tube is also susceptible, at least theoretically,  to buckling into an axially symmetric periodic mode \citep{sh1972, ho1979a, ho1979b, ch1997} and the limiting point instability \citep{al1971, kh2007, ren2011, hnh2015}. These two types of buckling/instability were not thought to be relevant to localised bulging for a long time and it was only recently that their connections had become clear \citep{fpl2008, fl2016, yf2022}.  In particular, it is now known that the bifurcation condition for localized bulging coincides with the condition for the limiting point instability in the case of free ends although this connection is lost in the case of fixed ends. Recognising that localised bulging is a \lq\lq zero wavenumber" bifurcation phenomenon, recent studies have examined the weakly nonlinear near-critical behaviour \citep{ylf2020}, and the effects of rotation, fibre-reinforcement, tethering and multi-layering on bulge initiation  \citep{wf2017, vd2017, wf2018, lya2019, yla2019}. The methodology developed has also been extended to study surface tension induced necking \citep{fjg2021, ef2021a, ef2021b, ef2021c}. The commonly used approach of modelling localised bulging/necking as a \lq\lq finite wavenumber" bifurcation phenomenon has been appraised in a recent study \citep{wf2021}.

The entire bulging process associated with the free-ends case under volume control is now fully understood. A localised bulge will initiate at a pressure value determined by the bifurcation condition. It will then grow in radius, accompanied by continuing pressure drops, until the radius at the centre of the bulge has effectively reached a maximum, after which the bulge will propagate axially at effectively a constant pressure (neither the maximum radius nor the constant pressure can be reached exactly in a tube of finite length no matter how long it is). This constant pressure and the associated maximum of radius are determined by Maxwell's equal area rule \citep{yi1977, ch1984}. In contrast, the propagation stage in the case of fixed ends does not seem to have been fully understood although Euler buckling due to unloading near the ends has been observed as a dominant feature experimentally and in numerical simulations \citep{pgl2006, takla2018, ddj2019, hhp2021}.
It was shown in  \citet{wg2019} that the pressure increases monotonically during the propagation stage, but it is not yet clear how this variation can be determined analytically. This question is addressed in the current paper. In particular, it is shown that at each propagation pressure the \lq\lq two phase" deformation is determined by an equal area rule in the force versus axial stretch diagram.

The rest of this paper is organised into five sections as follows. After summarising the bifurcation condition for localised bulging in the next section, we discuss predictions of three commonly used material models for localised bulging. It is shown that whereas the Gent and Gent-Gent material models predict the existence of a maximum pre-stretch above which localised bulging becomes impossible with fixed ends, the Ogden material model predicts otherwise. We then present experimental results for a typical commercially available latex rubber tube and confirm the existence of the above-mentioned maximum pre-stretch. In Section 4, we describe analytically the right ascending branch in the pressure versus bulging amplitude diagram and show that for different values of pre-stretch and tube lengths, this branch always converges to a single asymptote that we determine analytically. In Section 5, we highlight the fact that when the bifurcation curve is presented in the pressure/pre-stretch plane, the pressure has a minimum that is attained near the maximum pre-stretch mentioned above. The \lq\lq two-phase" solution in this parameter regime is determined analytically. The paper is concluded in Section 6 with a summary and some additional interpretations.

\section{Problem formulation}
\setcounter{equation}{0}
\setcounter{figure}{0}
We consider a hyperelastic cylindrical tube that is first subjected to an axial pre-stretch, denoted by $\lambda_z$, and then inflated by an internal pressure, denoted by $P$. We focus on the fixed-ends case whereby the length of the tube is fixed once the pre-stretch has been applied. The undeformed inner and outer radii are denoted by $A$ and $B$, respectively, and their deformed counterparts are $a$ and $b$.  In terms of cylindrical coordinates, the deformation is given by
\be r^2=\lambda_z^{-1} (R^2-A^2) +a^2, \;\;\;\; \theta =\Theta, \;\;\;\; z=\lambda_z Z, \la{rsz} \en
where  $(R, \Theta, Z)$ and $(r, \theta, z)$ are the  coordinates in the undeformed and deformed configurations, respectively.

With incompressibility taken into account, the three principal stretches are given by
$$ \lambda_1 \equiv \lambda =\frac{r}{R}, \;\;\;\; \lambda_2=\lambda_z, \;\;\;\; \lambda_3=\frac{dr}{dR}=1/(\lambda_1 \lambda_2), $$
where we have identified the indices $1, 2, 3$ with the $\theta$-, $z$-, and $r$-directions, respectively.

We assume that the constitutive behavior of the tube is described by a strain-energy function $W(\lambda_1, \lambda_2, \lambda_3)$. In terms of the reduced strain-energy function $w$ defined by
\be w(\lambda, \lambda_z)=W(\lambda, \lambda_z, \lambda^{-1} \lambda_z^{-1}), \la{reducedW} \en
the internal pressure is given by \citep{ho1979b}
\be
P= \int^{\lambda_a}_{\lambda_b} \frac{w_1}{\lambda^2 \lambda_z -1} d \lambda, \la{pressure} \en
where  $w_1=\paa w/\paa \lambda$, and the two limits $\lambda_a $ and $\lambda_b$ are defined by
$$ \lambda_a=\frac{a}{A}, \;\;\;\; \lambda_b =\frac{b}{B}, $$
and are related to each other by the volume-preserving condition
\be \lambda_a^2 \lambda_z-1=\frac{B^2}{A^2} (\lambda_b^2 \lambda_z-1). \la{ab} \en
The three principal Cauchy stresses are
\be
\sigma_{ii}=\lambda_i \frac{\paa W}{\paa \lambda_i}-\bar{p}, \;\;\;\; \hbox{no summation on}\, i, \la{stress} \en
where  $\bar{p}$ is the pressure associated with the constraint of incompressibility.

The resultant axial force at any cross section is independent of $Z$ and is given by \citep{ho1979b}
\be
N =2 \pi \int^b_a \sigma_{22} r dr -\pi a^2 P=\pi A^2 (\lambda_a^2 \lambda_z -1) \int^{\lambda_a}_{\lambda_b} \frac{2 \lambda_z w_2-\lambda w_1}{(\lambda^2\lambda_z-1)^2} \lambda d\lambda, \la{F} \en
where $w_2=\paa w/\paa \lambda_z$. In both \rr{pressure} and \rr{F} the $\lambda_b$ can be eliminated using the condition \rr{ab}.

It was shown numerically in \citet{fl2016} that the bifurcation condition for localised bulging is given by
 \be
 \Omega(\lambda_a, \lambda_z) \equiv \frac{\paa P}{\paa \lambda_a} \frac{\paa N}{\paa \lambda_z} -\frac{\paa P}{\paa \lambda_z} \frac{\paa N}{\paa \lambda_a}=0, \la{jacobian} \en
 which states that the Jacobian of the functions $P$ and $N$ vanishes. An analytical derivation of this condition has recently been given by   \citet{yf2022}.

From now on, we assume that the pressure has been scaled by $\mu H/R_m$ and the axial force by $2 \pi \mu R H$, where $\mu$ is the ground-state shear modulus, $R_m=(A+B)/2$ and $H=B-A$. When the tube is modeled by the membrane theory, we use $\lambda_a$ to denote the azimuthal stretch at the mid-surface $R=R_m$ to avoid introducing an extra notation.

\section{Bulge initiation based on different material models}

We consider the Ogden \citep{og1972}, Gent \citep{ge1996}, and Gent-Gent \citep{pucci2002} material models given by
\be W=\sum_{r=1}^{3} \mu_r
(\lambda_1^{\alpha_r}+\lambda_2^{\alpha_r}+\lambda_3^{\alpha_r}
-3)/\alpha_r, \la{ogdenmat} \en
\be W=-\frac{1}{2} \mu J_m \ln
(1-\frac{I_1-3}{J_m}),   \la{gentmat} \en
\be W=-\frac{\mu_0}{2} J_m \ln(1-\frac{I_1-3}{J_m})+C_2 \ln (\frac{I_2}{3}), \la{gent-gent} \en
respectively, where $\mu_0$, $\mu$, $J_m$, $C_2$, and $(\alpha_r, \mu_r)$ ($r=1, 2, 3$) are material constants, and $I_1$ and $I_2$ are the first and second principal invariants of the Cauchy-Green deformation tensors. Corresponding to \rr{gent-gent}, the ground state shear modulus is given by $\mu=\mu_0+2 C_2/3$.
In our previous study, \citet{wg2019}, we conducted a series of experiments on localised bulging of inflated rubber tubes in order to verify various theoretical predictions. The rubber material was characterised using uni-axial tension, pure shear and biaxial tension. Based on the considerations in \citet{zw2018}, we adopted the
Gent-Gent material model and obtained the values
\be J_m=88.43, \;\;\; \mu_0=0.2853\, {\rm MPa},\;\;\;\; C_2=0.1898\, {\rm MPa} \la{Gent-Gent1} \en
using data fitting. The corresponding shear modulus is $\mu=0.4118\, {\rm MPa}$.
For our current study, we also fit the Gent material model to the same experimental data provided by  \citet{wg2019}. We find that
\be
J_m=97.2, \;\;\;\; \mu=0.32\, {\rm MPa}. \la{Gent} \en
In adopting Ogden's material model, based on the discussions by  \citet{oss2004}, we decide to use the same values for $\alpha_r$ and $\mu_r/\mu$ as given by   \citet{og1972}, that is
$$
 \alpha_1=1.3, \;\;\;\; \alpha_2=5.0, \;\;\;\;\alpha_3=-2.0, $$ \be
\quad \mu_1=1.491 \mu, \;\;\;\; \mu_2=0.003 \mu , \;\;\;\; \mu_3=-0.024 \mu, \la{Og} \en
but the shear modulus $\mu$ is obtained by fitting the model to the same experimental data provided by  \citet{wg2019}. This gives
$\mu=0.4627\, {\rm MPa}$ which is slightly higher than the value of $0.414$ given by   \citet{og1972}.

We shall present our results mostly under the membrane assumption. The reasons are two-fold. Firstly, adopting the membrane assumption will enable us to explain the methodology without having to deal with the heavy algebra that is necessary when a tube of arbitrary wall thickness is considered.
Secondly, it is known that results obtained under the membrane assumption can be valid for values of $H/R_m$ as large as $0.67$ \citep{fl2016}.

Under the membrane assumption, the scaled pressure and axial force are free of integrals and have the simple expressions
\be
P=\tilde{P}(\lambda_a, \lambda_z)\equiv\frac{w_1}{\lambda_a \lambda_z}, \;\;\;\; N=\tilde{N}(\lambda_a, \lambda_z) \equiv w_2-\frac{\lambda_a w_1}{ 2 \lambda_z}, \la{1.1} \en
where the "$\equiv$" sign  define the two functions  $\tilde{P}(\lambda_a, \lambda_z)$ and $\tilde{N}(\lambda_a, \lambda_z)$.
The bifurcation condition \rr{jacobian} then reduces to
\be \lambda_a (w_1-\lambda_z w_{12})^2+\lambda_z^2 w_{22} (w_1-\lambda_a w_{11})=0. \la{bifmem} \en

\begin{figure}[ht]
\begin{center}
\begin{tabular}{ccc}
 \includegraphics[width=.4\textwidth]{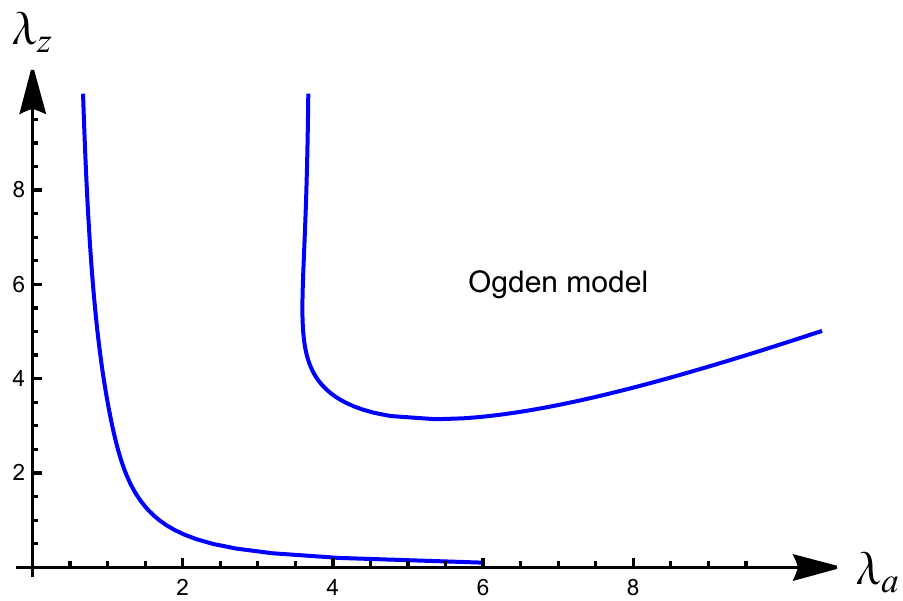} &  &\includegraphics[width=.4\textwidth]{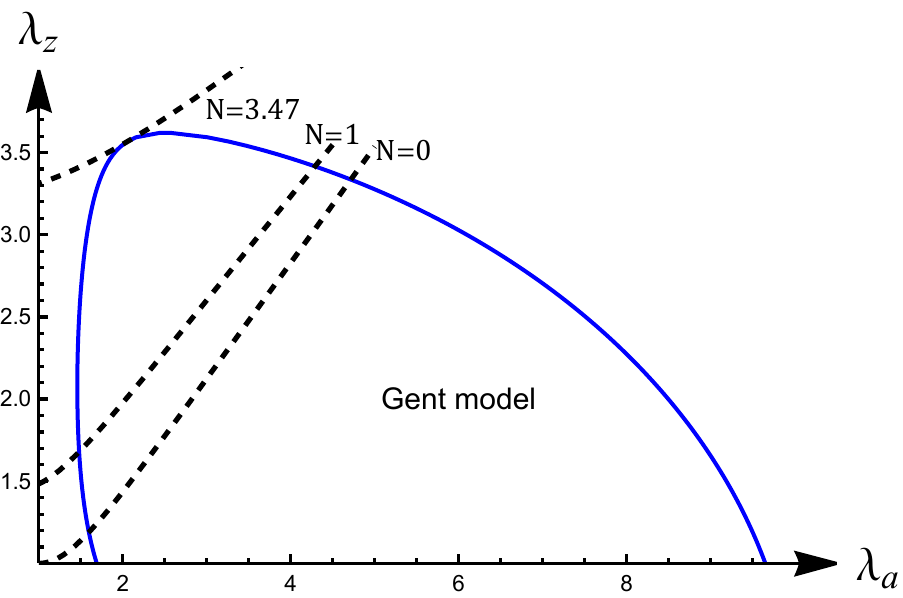} \\
 (a) & &(b)
\end{tabular}
\end{center}
\caption{Bifurcation condition corresponding to (a) Ogden material model, and (b) the Gent material model. The Ogden model allows localised bulging to occur at any pre-stretch value, whereas under the Gent material model localised bulging is only possible if the pre-stretch is less than $3.62$.}
\label{newfig1}
\end{figure}

\begin{figure}[ht]
\begin{center}
\begin{tabular}{ccc}
 \includegraphics[width=.4\textwidth]{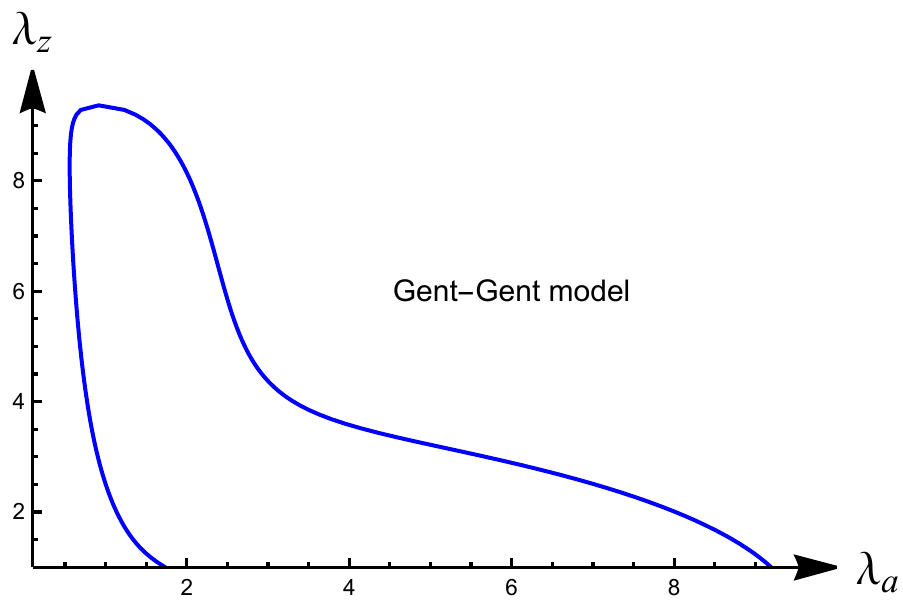}  & &\includegraphics[width=.4\textwidth]{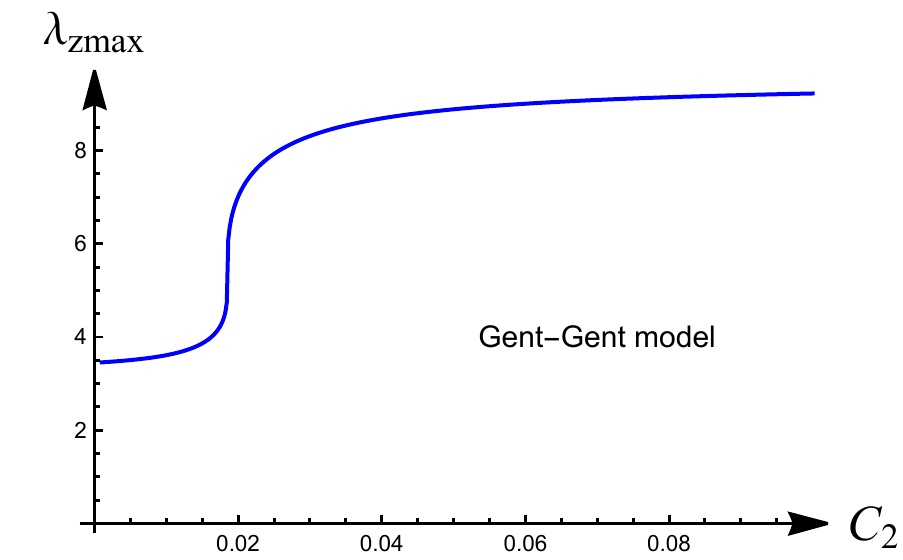}\\
 (a) & &(b)
\end{tabular}
\end{center}
\caption{(a) Bifurcation condition corresponding to the Gent-Gent material model, showing that localised becomes impossible only when the pre-stretch becomes as large as $9.37$.  (b) Dependence on $C_2$ of the maximum pre-stretch $\lambda_{z{\rm max}}$ above which localised bulging becomes impossible. The maximum pre-stretch $\lambda_{z{\rm max}}$ increases rapidly around $C_2 =0.02$.}
\label{newfig2}
\end{figure}

Figs~\ref{newfig1}(a, b) and \ref{newfig2}(a) display the bifurcation condition \rr{bifmem} for the three material models given by \rr{ogdenmat}--\rr{gent-gent}. For each material model, the initiation pressure for localised bulging can be determined graphically as follows. If inflation is carried out with fixed axial force $N$, say
$N=1$, then the loading path in the $(\lambda_a, \lambda_z)$-plane is a curve represented by $\tilde{N}(\lambda_a, \lambda_z)=1$ which is shown in Figs~\ref{newfig1}(b). The first (i.e. bottom left) intersection with the bifurcation curve gives the bifurcation values of $\lambda_a$ and $\lambda_z$ for localised bulging. In the case of fixed ends with a specified pre-stretch,  the loading path is a horizontal line in the $(\lambda_a, \lambda_z)$-plane and the first intersection with the bifurcation curve gives the bifurcation value of $\lambda_a$ for localised bulging. In either case, the initiation pressure is computed with the use of \rr{1.1}$_1$. It is found that for values of $N$ and $\lambda_z$ in a sufficiently small neighbourhood of $N=0$ and $\lambda_z=1$, the initiation pressures predicted by the three material models are very close to each other. For instance, in the case of fixed axial force with $N=1$, the initiation pressures predicted by the Ogden, Gent, Gent-Gent models are 0.59, 0.56, and 0.61, respectively, with the predictions by the Ogden and Gent-Gent models differing by only 3.3\%. However, their predictions begin to diverge for larger values of $N$ and $\lambda_z$. Most importantly, there exists the qualitative difference that whereas both the Gent and Gent-Gent models predict the existence of a maximum axial force or pre-stretch beyond which the loading path has no intersections with the bifurcation curve (and hence localised bulging becomes impossible), the Ogden model predicts otherwise due to the fact that the associated bifurcation curve extends to infinity in the $\lambda_z$ direction. Also, the maximum pre-stretches predicted by the Gent and Gent-Gent models are $3.62$ and $9.37$, respectively, which are quite different. It is then of interest to investigate whether such a pre-stretch maximum exists or not for commonly used rubber materials, and if it does, whether it is accurately predicted by the Gent and Gent-Gent material models. Since the Gent-Gent model reduces to the Gent model when $C_2=0$, to understand the significant difference in the above predictions, we have shown in Figs~\ref{newfig2}(b) the dependence of the maximum pre-stretch $\lambda_{z{\rm max}}$ on the parameter $C_2$ with the other two parameters $J_m$ and $\mu$ fixed at the values given by \rr{Gent-Gent1}. It is seen that $\lambda_{z{\rm max}}$ experiences a jump as $C_2$ crosses a value approximately equal to $0.02$.

Motivated by the above observations, we decided to conduct additional experiments for the case of fixed ends with large pre-stretches. We used tubes made of natural latex rubber that are sold on the market as exercise bands and catheters. The tubes used all have inner radius $3$ mm,  outer radius  $4.5$ mm,  and the effective experimental length is $200$ mm. The experimental set-up is the same as that described in our earlier paper, \citet{wg2019}. The measurement system mainly includes air compressor, pressure regulating valve, throttle valve, tension sensor, pressure sensor, stretching module, data acquisition card. The tension and pressure sensors are used to convert analog signal voltage output to the data acquisition card. With a volume of $8$ liters, the air compressor can provide sufficient pressure. The pressure regulating valve has a set range of $0-0.2$ MPa,
and the tension sensor has a range of $0-100$ N.
The pressure sensor has a range  $0-0.2$ MPa with an accuracy of 0.25\%FS.
Before each experiment, the sensors were calibrated to determine the relationship between voltage, tension, and pressure. The pre-stretch is controlled by a motor screw system, and the displacement range is $0-1.5$ m. We measured the internal pressure at one end of the tube. We assume that the pressure in the tube is rapidly homogenized and so the pressure is uniform everywhere.

Before each experiment, the tube was repeatedly stretched 3 times, each time to twice the original length of the tube. The tube was subsequently installed on the experimental device, and the distance between the two fixtures was taken as the initial length $L$. After the tube has been stretched to length $l$ so that $\lambda_z =l /L$, the two ends are fixed and inflation begins.

\begin{figure*}[ht]
\begin{center}
\begin{tabular}{ccc}
\includegraphics[width=.45\textwidth]{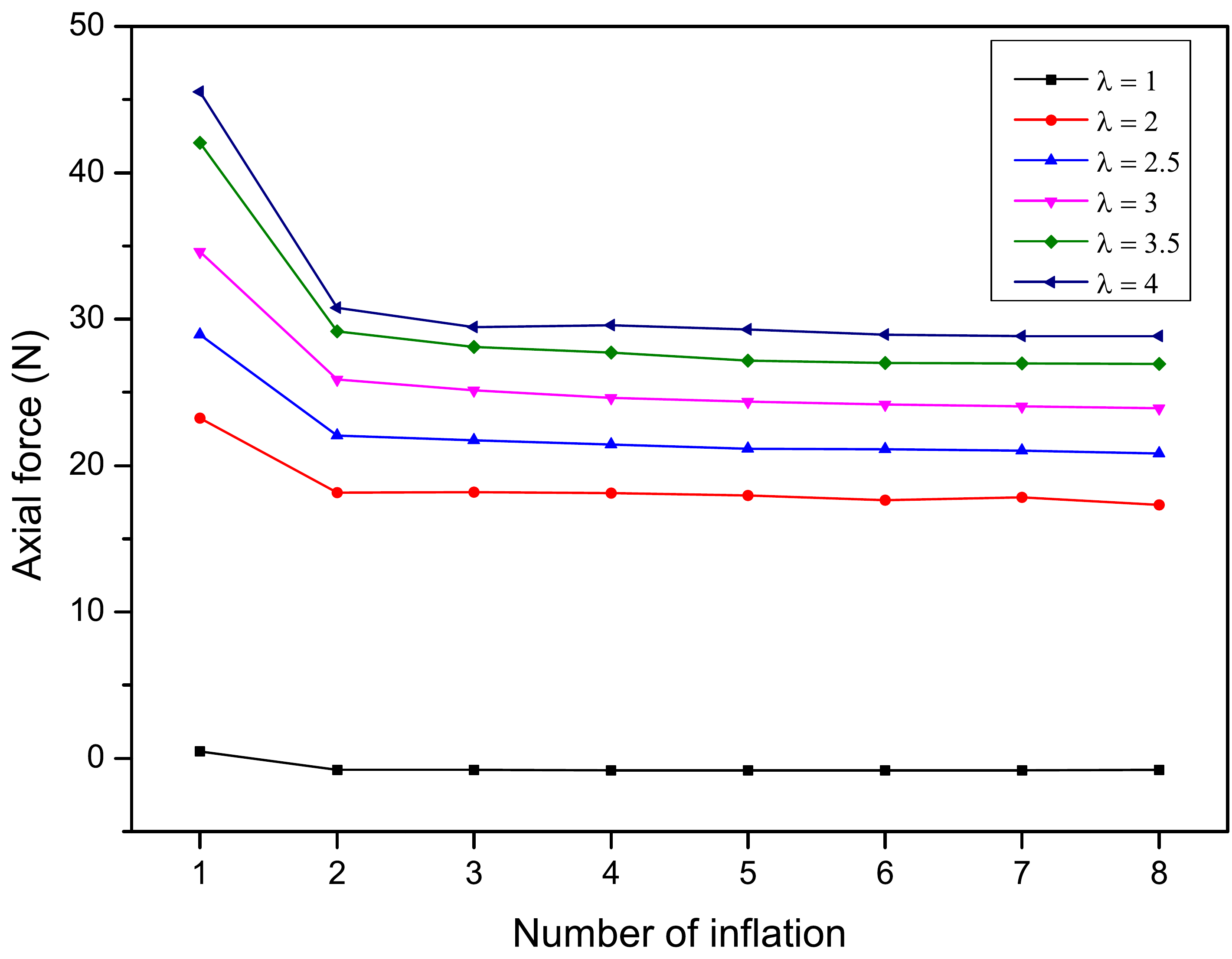} & & \includegraphics[width=.45\textwidth]{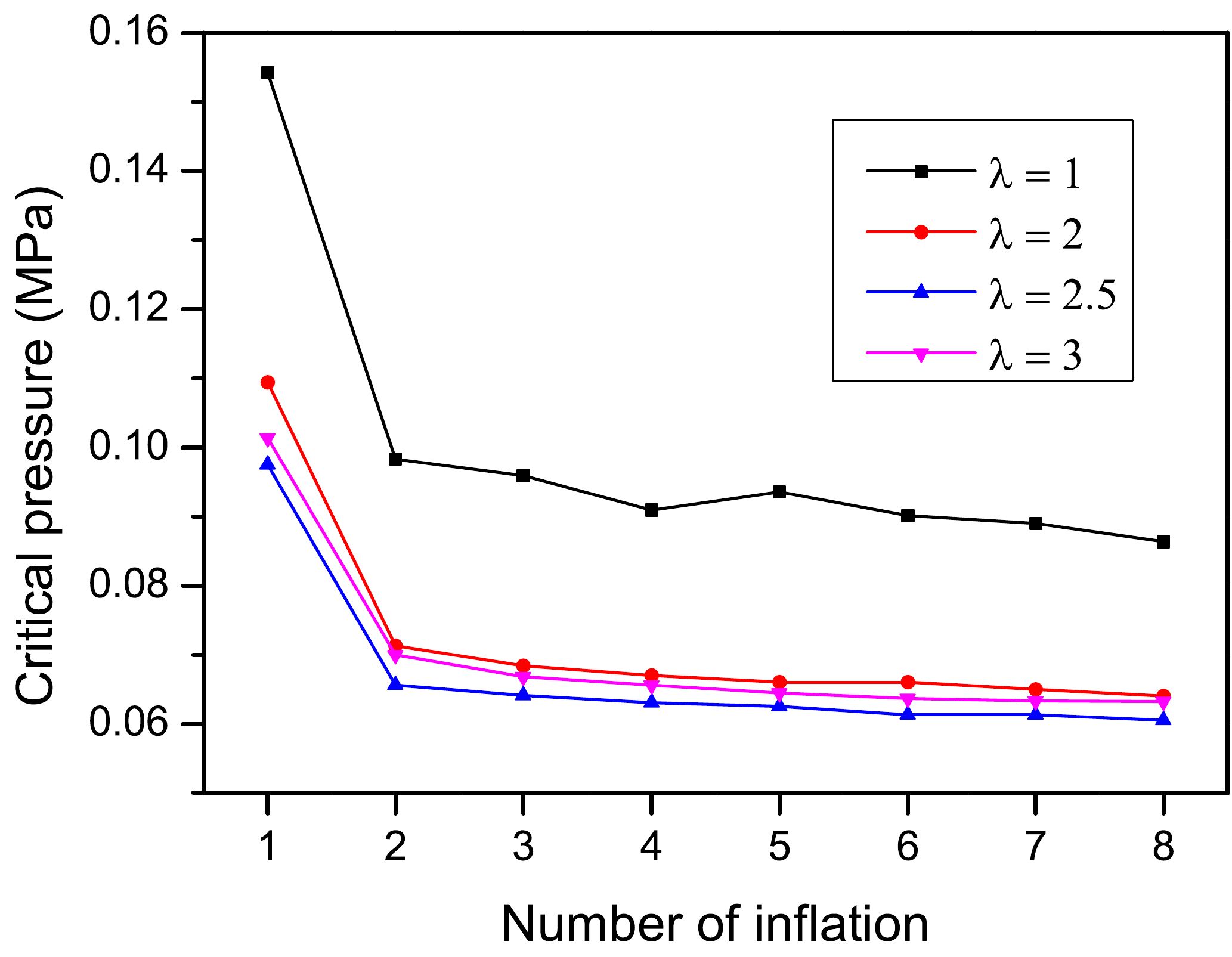}  \\ \\
(a) & & (b)
\end{tabular}
\end{center}
\caption{\small{ Dependence of (a) axial force and (b) critical pressure for localised bulging on the number of inflations under different values of the pre-stretch $\lambda_z$}.}
\label{fig045}
\end{figure*}

%
%
%

Fig.\,\ref{fig045}(a) displays the axial force required to maintain the prescribed axial pre-stretch before the start of each inflation. It shows that the above mentioned pre-stretching alone is not sufficient to remove Mullins effect. The latter only disappears after two full inflations. It can be seen that the larger the pre-stretch is, the greater the drop in the axial force is due to the first inflation. When the pre-stretch is 4, the initial tension decreases from 45.5 N to 30.8 N after the first inflation, a decrease of  32.4\%. When the pre-stretch is 3.5, 3, 2.5 and 2, respectively, the corresponding reductions in the initial tension are 30.6\%, 25.2\%, 23.8\% and 21.9\%, respectively. When there is no pre-stretch $\lambda_z=1$, the axial tension is essentially 0 although after the first full inflation, the axial force is negative, which is due to the residual deformation of the tube caused by the inflation.

Fig.\,\ref{fig045}(b) shows that the initiation (critical) pressure for localised bulging becomes independent of the number of inflations after two full inflations, again  verifying the above observation that Mullins effect can be eliminated by two full inflations (a full inflation is one in which the bulge has appeared and propagated to the ends of the tube).  When the pre-stretch is $1$, the critical pressure of the tube was reduced from 0.154 to 0.098 by 36.3\% between the first two inflations, and decreased by only 2.4\% from the second to the third inflations. After that, the critical pressure remained unchanged with the increase of inflation times.

\begin{figure*}[ht]
\begin{center}
\begin{tabular}{ccc}
\includegraphics[width=.45\textwidth]{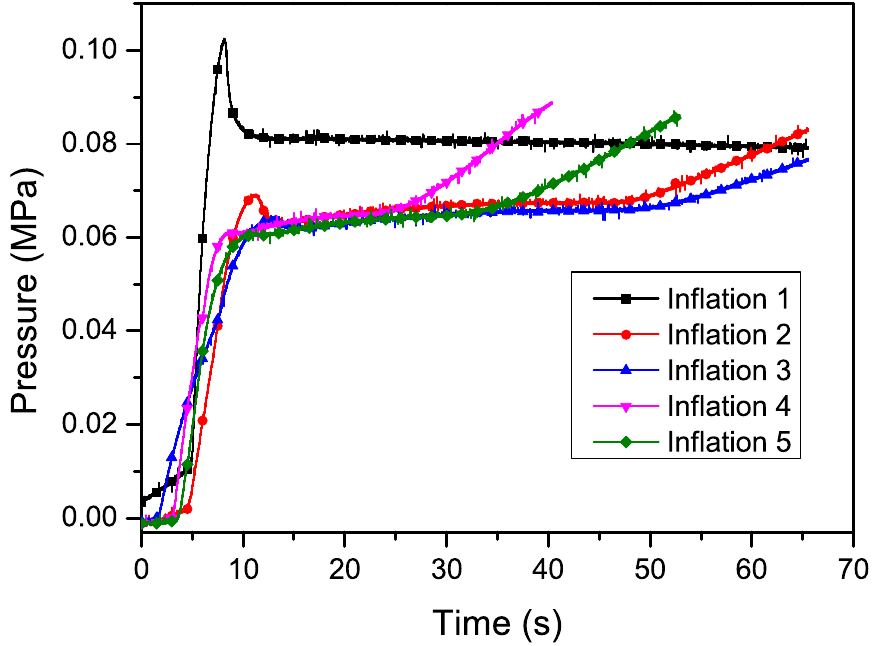} & &\includegraphics[width=.45\textwidth]{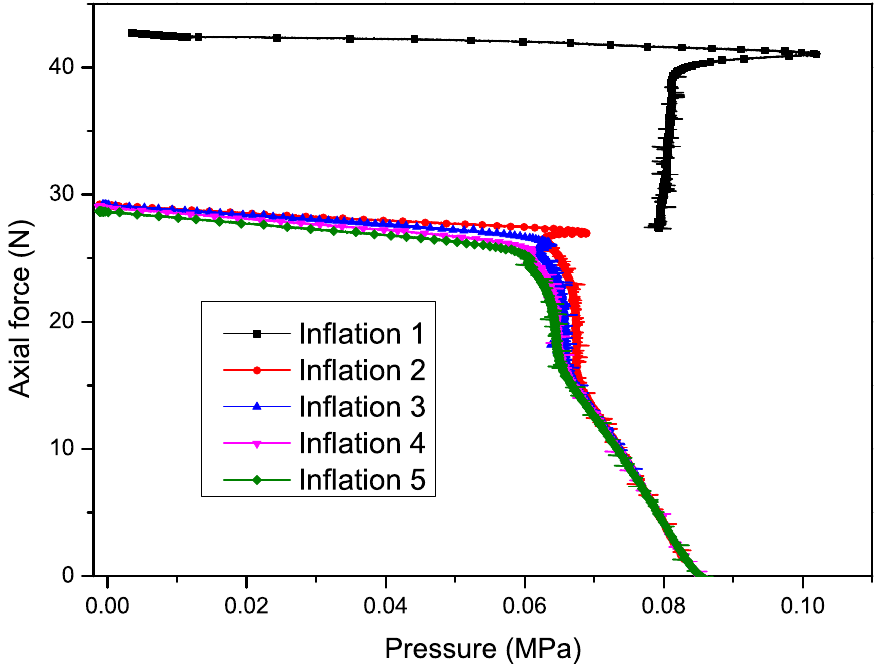}  \\ \\
(a) & & (b)
\end{tabular}
\end{center}
\caption{\small{  (a) The recorded history of internal pressure and time of 5 inflations, $\lambda_z=4$; (b) The curve of internal pressure and axial force of different inflation. $\lambda_z=4$.}}
\label{fig067}
\end{figure*}

%

The effect of a pre-stretch in the range $1.4 \le \lambda_z \le 2.5$ has been studied in  \citet{wg2019}. When the pre-stretch is larger, localised bulging exhibits new characteristics. Fig. \ref{fig067}(a) shows the pressure variation with respect to time in five inflations of the same tube that has been subjected to a pre-stretch of $4$. In the first inflation, localised bulging occurs when the pressure reaches 0.105 MPa. Localised bulging still occurs in the second inflation, but disappears in subsequent inflations.

For the current case of fixed ends, the axial force keeps changing during the inflation process. Fig.\,\ref{fig067}(b) shows the relationship between the pressure and axial force when the pre-stretch is equal to $4$. It is seen that in the first two inflations when localised bulging takes place, the axial force decreases quite slowly in uniform inflation but drops rapidly during the propagation stage. In subsequent inflations, the relationship can clearly be divided into two distinct phases. The first phase corresponds to uniform inflation in which axial force decreases quite slowly, whereas in the second phase the axial force unloads rapidly as pressure is increased, culminating in Euler buckling (for thin-walled tubes) or axisymmetric periodic buckling (for thick-walled tubes).

Table 1 shows the critical (initiation) pressure in multiple inflations with a range of pre-stretch values. The same batch of tubes were used in all experiments, but different tubes were used for different pre-stretch values. It can be seen from Table 1 that when the pre-stretch is less than 3, localised bulging occurs in all inflations.   However, when the pre-stretch is $3.5$, $4$, or $4.5$, respectively, localised bulging does not occur starting from the 5th, 4th and the 3rd inflation, respectively. Based on these results, we may conclude that localised bulging becomes impossible when the pre-stretch is approximately as large as 4.5. Thus, the Gent material model slightly under-predicts this value, the Gent-Gent material model over-predicts this value, whereas the Ogden material model cannot predict this behaviour. Interestingly, in contrast with the observation made by   \citet{hhp2021} that rupture occurs when the pre-stretch is equal to or larger than 3, we have not observed rupture before the bulge (if it occurs) has propagated to the tube ends. The only explanation that we may offer is  that the particular latex rubber tubes that we used are less prone to rupture.

%

\begin{table}[tp]
\centering
\setlength{\belowcaptionskip}{10pt}%
\caption{Critical pressure (unit: MPa) in different inflations and under different pre-stretches}
\begin{tabular}{cccccccc}  
\hline
 $\lambda_z=$  & 1  & 2 & 2.5 & 3 & 3.5 & 4 & 4.5\\ \hline  
inflation 1  &0.1542&0.1094&	0.0975 &	0.1014 &	0.1005&	0.1128&	0.136\\          
inflation 2 &0.0983&0.0713&	0.0656	&0.07&	0.0732&	0.075&0.0825 \\        
inflation 3 &0.0959&	0.0684	&0.0641&0.0668&	0.0695&	0.0681&x \\
inflation 4&0.0909	&0.067&	0.0631&	0.0656&	0.0662	&x	&x \\
inflation 5&0.0936	&0.066	&0.0625	&0.0645	&x&	x&	x \\
inflation 6&0.0901&	0.066&	0.0613&	0.0637	&x	&x	&x \\
inflation 7&0.089&	0.065	&0.0613	&0.0633	&x&	x	&x \\
inflation 8&0.0864&	0.064&	0.0605	&0.0632	&x	&x	&x \\ \hline
\end{tabular}
\end{table}

\section{Characterisation of the propagation stage}
Having established the fact that the Gent material model can qualitatively predict the existence of a limiting pre-stretch better than the Gent-Gent model or the Ogden model, we now characterise the propagation stage using the Gent material model.

\begin{figure*}[ht]
\begin{center}
\begin{tabular}{ccc}
\includegraphics[width=.45\textwidth]{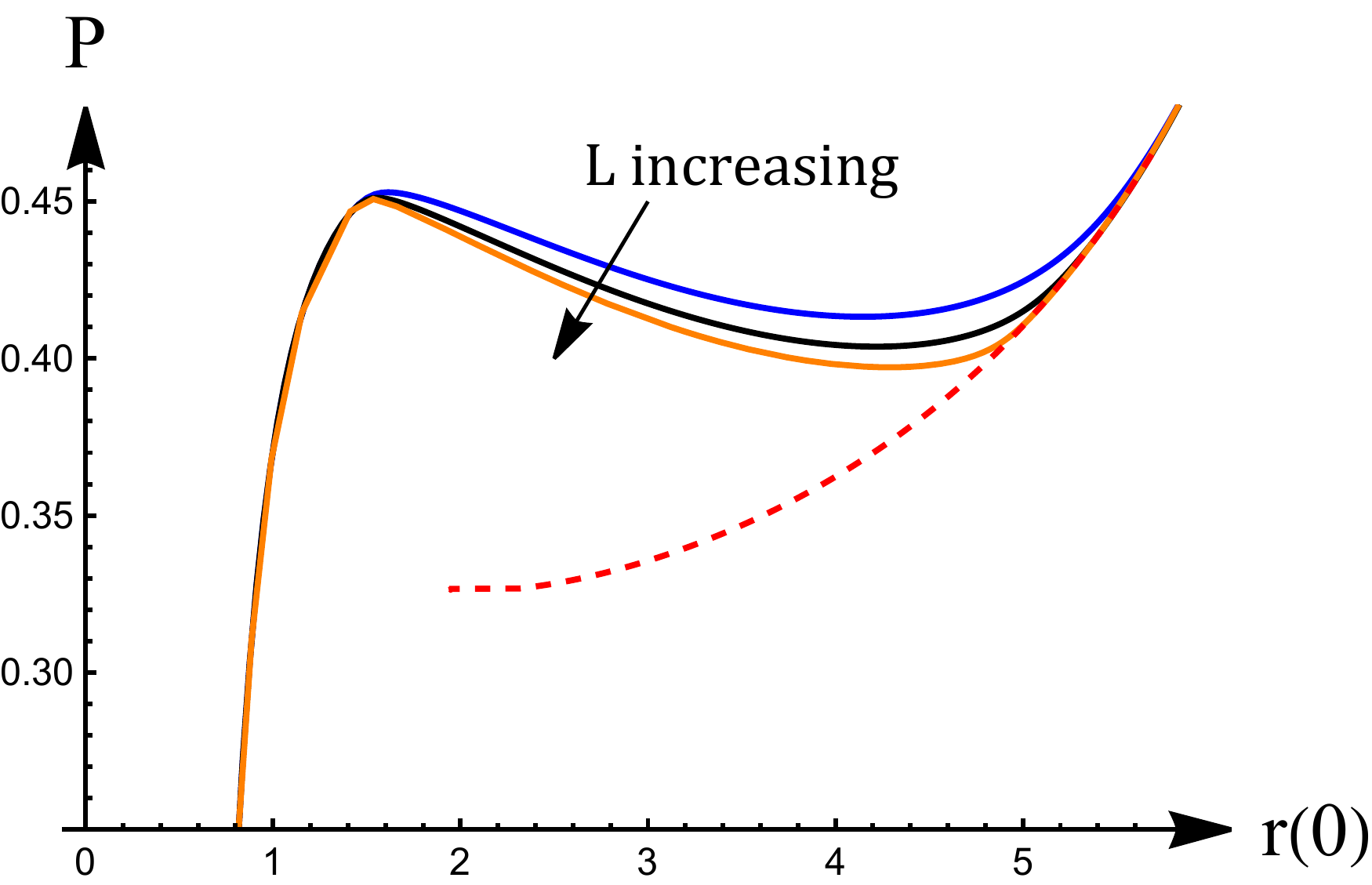} & & \includegraphics[width=.45\textwidth]{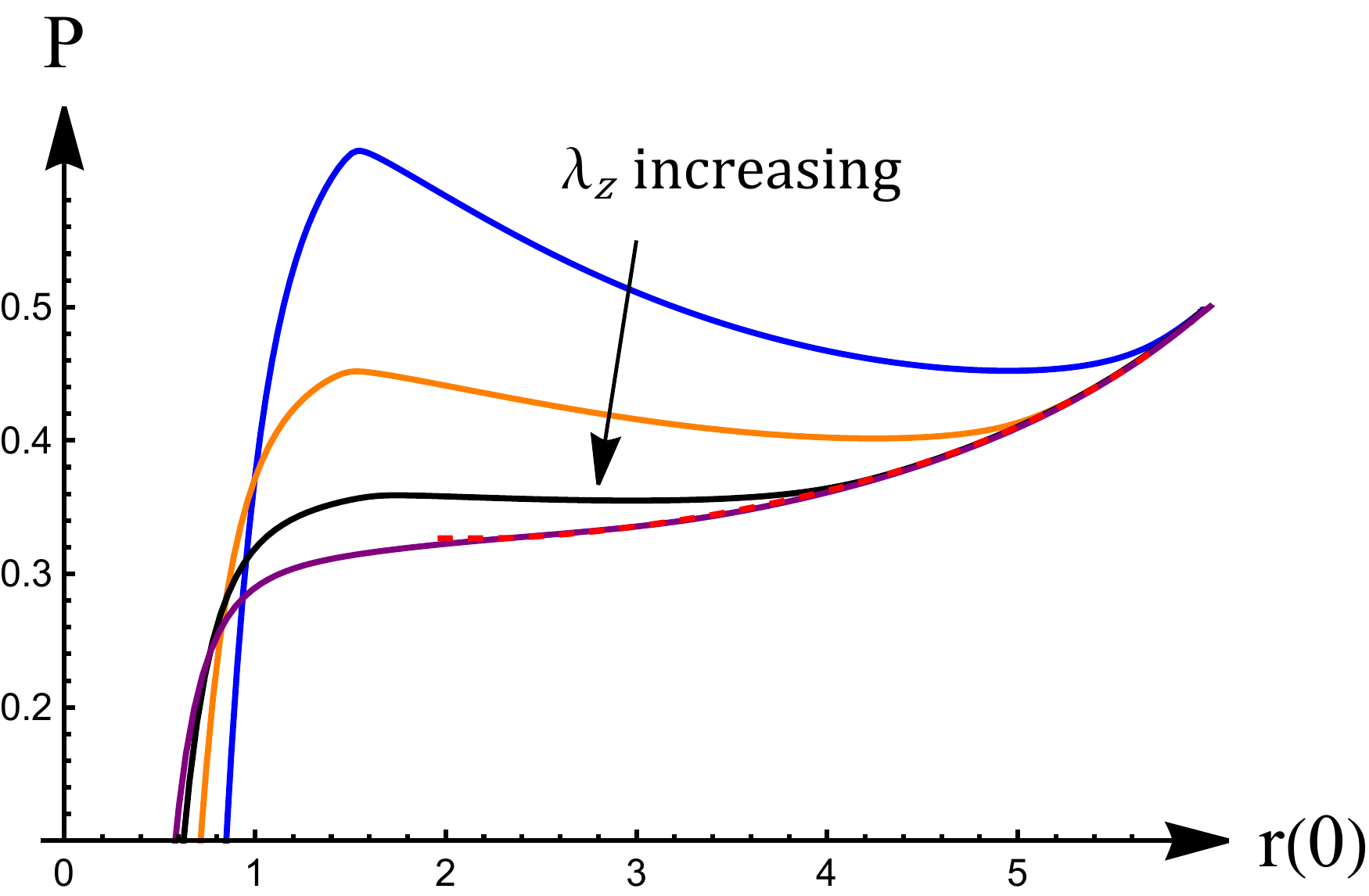}  \\ \\
(a) & & (b)
\end{tabular}
\end{center}
\caption{\small{ Dependence of $P$ on $r(0)$ (radius at $R=0$) when inflation is carried out with fixed ends with (a) $\lambda_z$ fixed at 2.2 and $L/R=30, 50$ and $90$,
respectively, and (b) $L/R$ fixed at 60 and $\lambda_z=1.5, 2.2$, $3$ and $3.5694$, respectively. In each plot, the dashed line corresponds to Maxwell's equal area rule in the $N$ versus axial stretch diagram in Fig.\ref{fig10ab}(a) at each fixed $P$. The lower end of the dashed line corresponds to the minimum pressure below which localised bulging become impossible under axial tension. The solid lines are Abaqus simulation results for tubes with $H/R=0.1$ and they all converge to the dashed line. It is when each solid line is sufficiently close to the dashed line that the bulge starts to propagate rapidly in the axial direction.}}
\label{fig6}
\end{figure*}

It was reported in   \citet{wg2019} that in the case of fixed ends, the pressure would increase to a maximum under uniform inflation, reach a minimum after the initiation of a bulge, and then increase monotonically. The latter ascending branch was not described analytically. In view of the results reported in Figure 13 of \citet{fjg2021} for an analogous problem, see also \citet{hb2014} and \citet{xb2017}, we anticipate that the right ascending branch mentioned above should approach a single asymptote that is independent of the
pre-stretch or the tube length. This is verified in Fig.~\ref{fig6} where we present our Abaqus simulation results for a variety of pre-stretch values and tube lengths. Such an asymptote does indeed exist, as shown in Fig.~\ref{fig6} in dashed line. This asymptote is determined as follows.

\begin{figure}[ht]
\begin{center}
\begin{tabular}{ccc}
 \includegraphics[width=.4\textwidth]{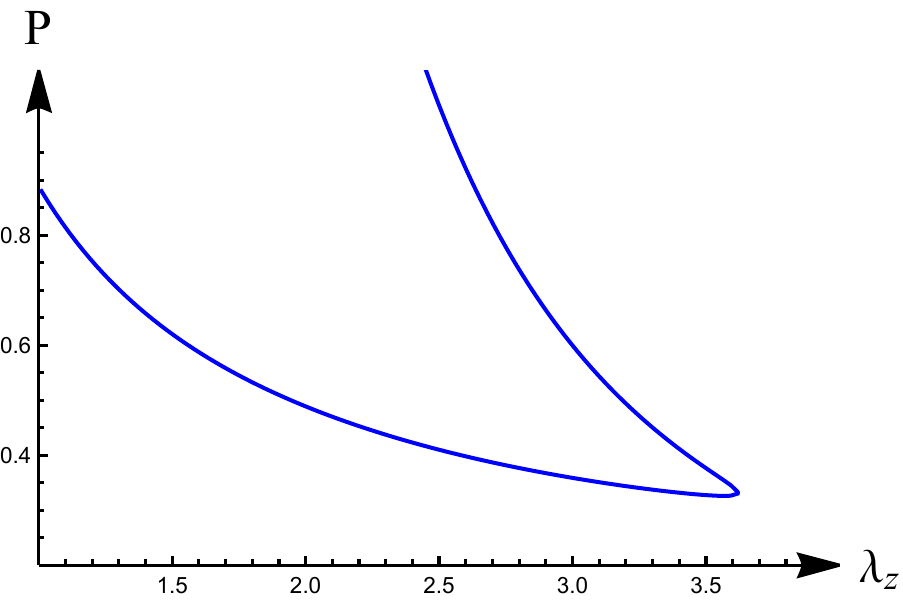} &  &\includegraphics[width=.4\textwidth]{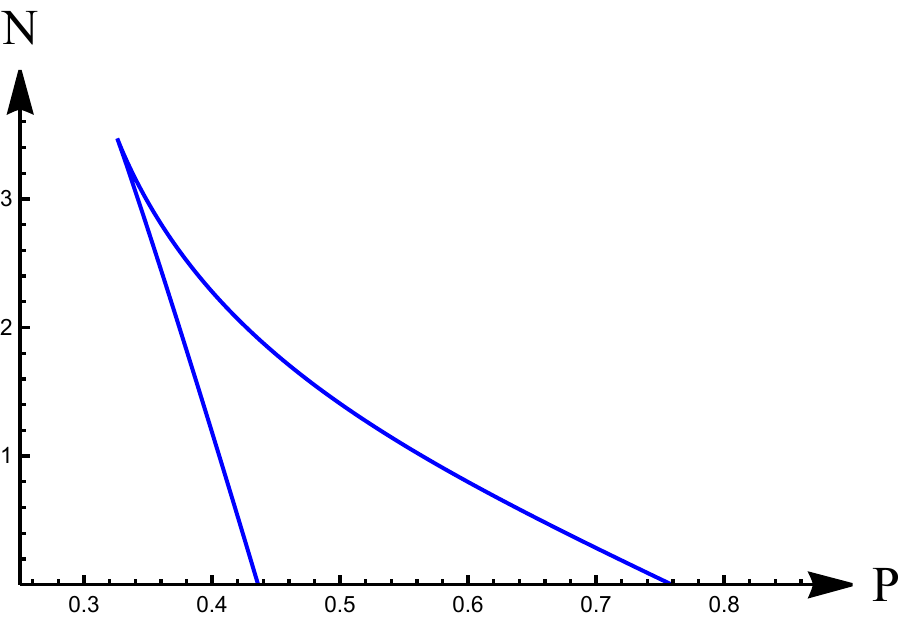}\\
 (a) & &(b)
\end{tabular}
\end{center}
\caption{(a) Bifurcation condition in the $\lambda_z-P$ plane. The pressure minimum $P=0.3259$ occurs at $\lambda_z=3.5694$ and $\lambda_a=2.0628$ where the loading curve
$\tilde{N}(\lambda_a, \lambda_z)=3.4712$ becomes tangent to the bifurcation curve in Fig.~\ref{newfig1}(b); see Fig.\,\ref{fig11}(a) for a close-up near the pressure minimum.  The maximum of
$\lambda_z$ is $3.6199$ and is attained when $P=0.3316$ and $\lambda_a=2.5050$.
 (b) Bifurcation condition in the $P-N$ plane. The minimum of $P$ is the same as in (a) and the corresponding $N$ is $3.4712$.}
\label{fig7ab}
\end{figure}

We refer to Fig.~\ref{newfig1}(b) and note that there exists a single loading curve in the case of fixed axial force, approximately given by $\tilde{N}(\lambda_a, \lambda_z)=3.4712$, that is tangent to the bifurcation curve. Localised bulging becomes impossible if $N>3.4712$. In a similar way, in the case of fixed ends, the horizontal line $\lambda_z=3.6199$ is tangent to the bifurcation curve, and localised bulging becomes impossible if $\lambda_z>3.6199$. These facts are made more transparent in Fig.~\ref{fig7ab}(a, b) where the bifurcation curve is presented in the $(P, \lambda_z)$-plane and $(P, N)$-plane, respectively. The curve in Fig.~\ref{fig7ab}(a) is obtained by expressing $\lambda_a$ as a function of $\lambda_{z}$ through the bifurcation condition, and then plot $P$ against $\lambda_{z}$. The curve in Fig.~\ref{fig7ab}(b) is obtained using parametric plotting by viewing $P$ and $N$ as functions of $\lambda_{z}$, with $\lambda_a$ determined by the bifurcation condition. The maximum of $N$ in Fig.~\ref{fig7ab}(b) is given by $N=3.4712$. This figure also provides the additional information that localised bulging is impossible under axial tension if $P$ is less than $0.3259$, the minimum value of $P$ in Fig.~\ref{fig7ab}(a) attained at $\lambda_z=3.5694$. This minimum also features in the force $N$ versus $\lambda_{z}$ diagrams to be discussed next.

\begin{figure*}[ht]
\begin{center}
\begin{tabular}{ccc}
\includegraphics[width=.35\textwidth]{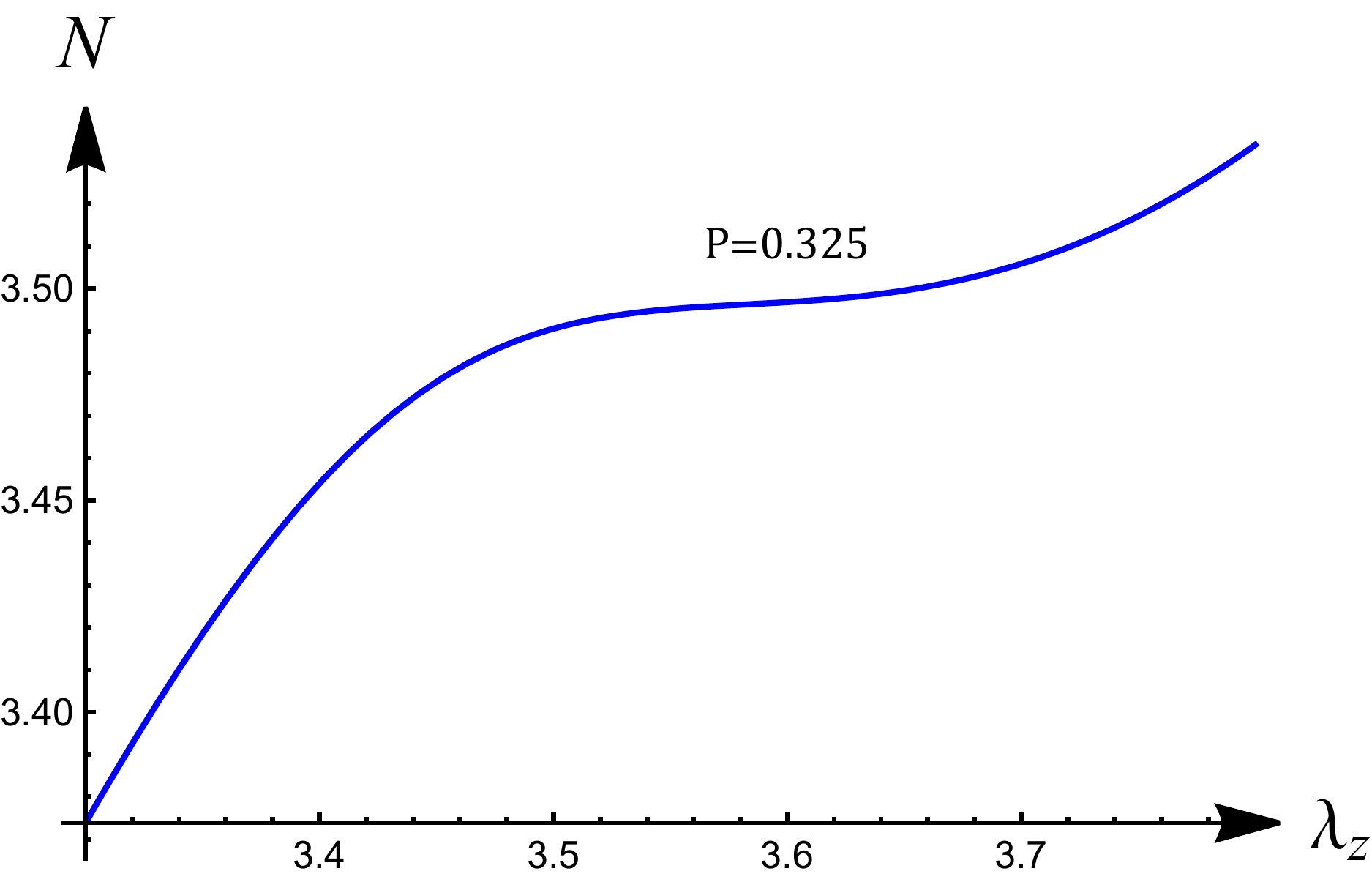} & & \includegraphics[width=.35\textwidth]{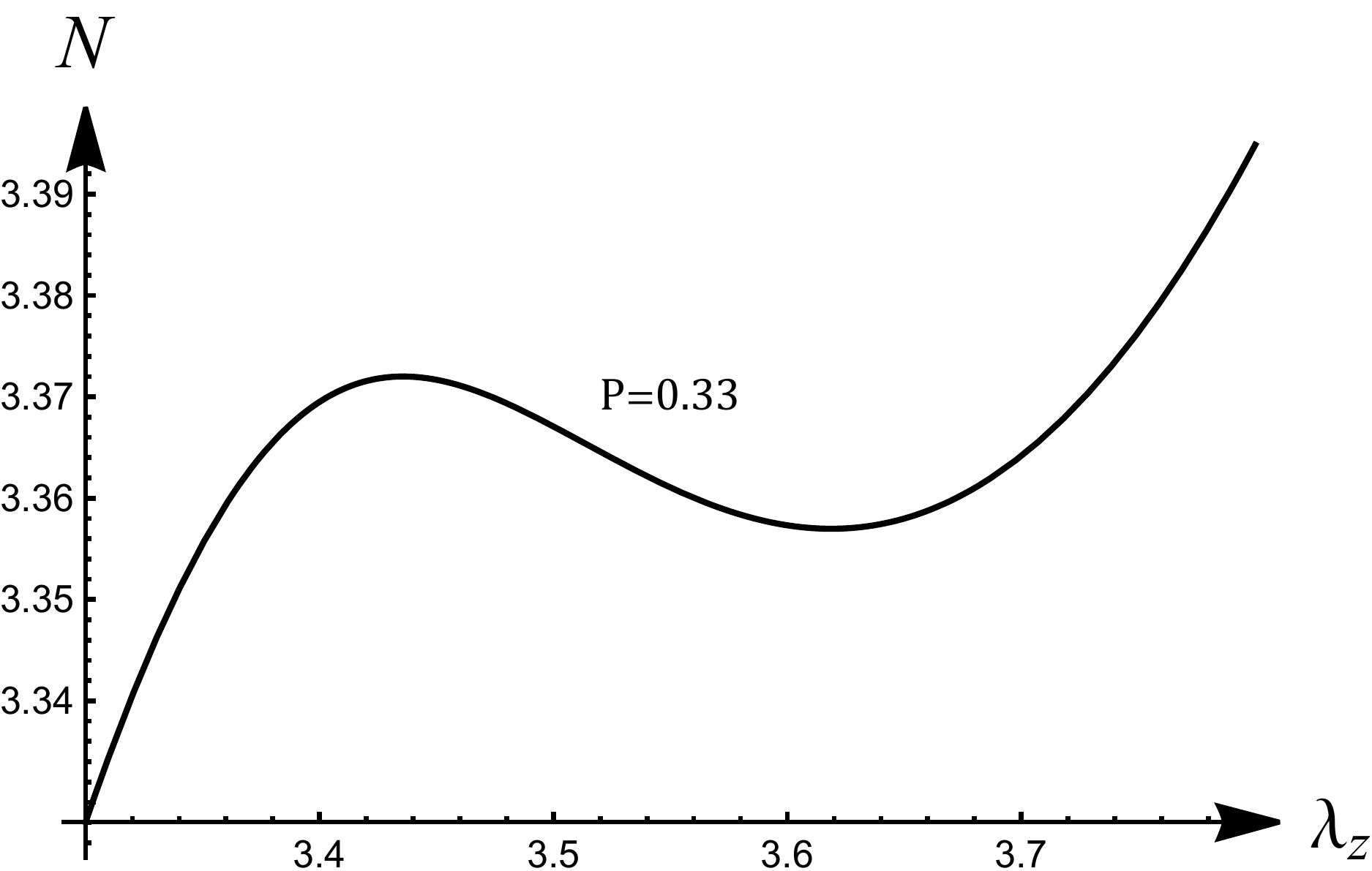} \\  \\
\includegraphics[width=.35\textwidth]{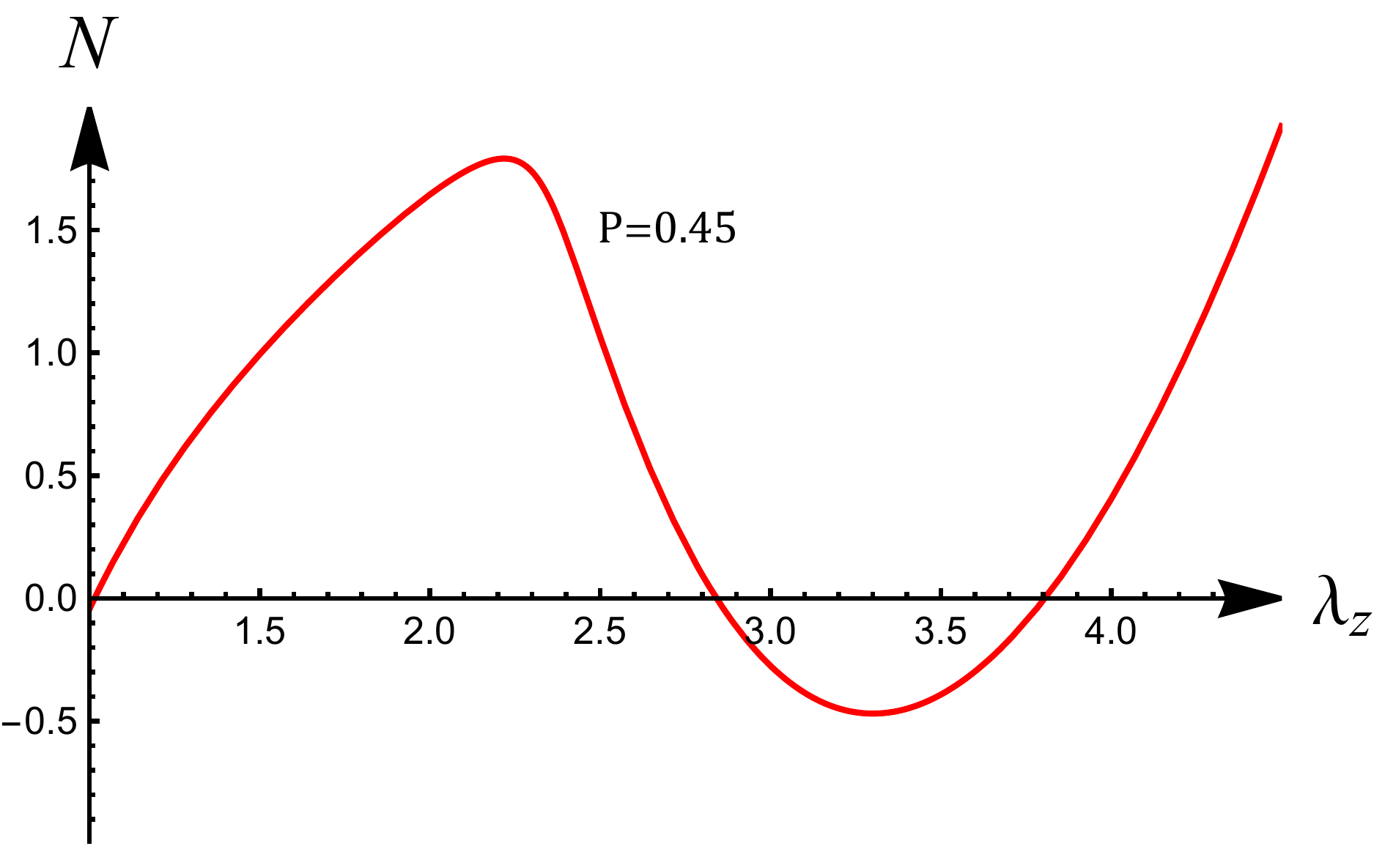} & &\includegraphics[width=.35\textwidth]{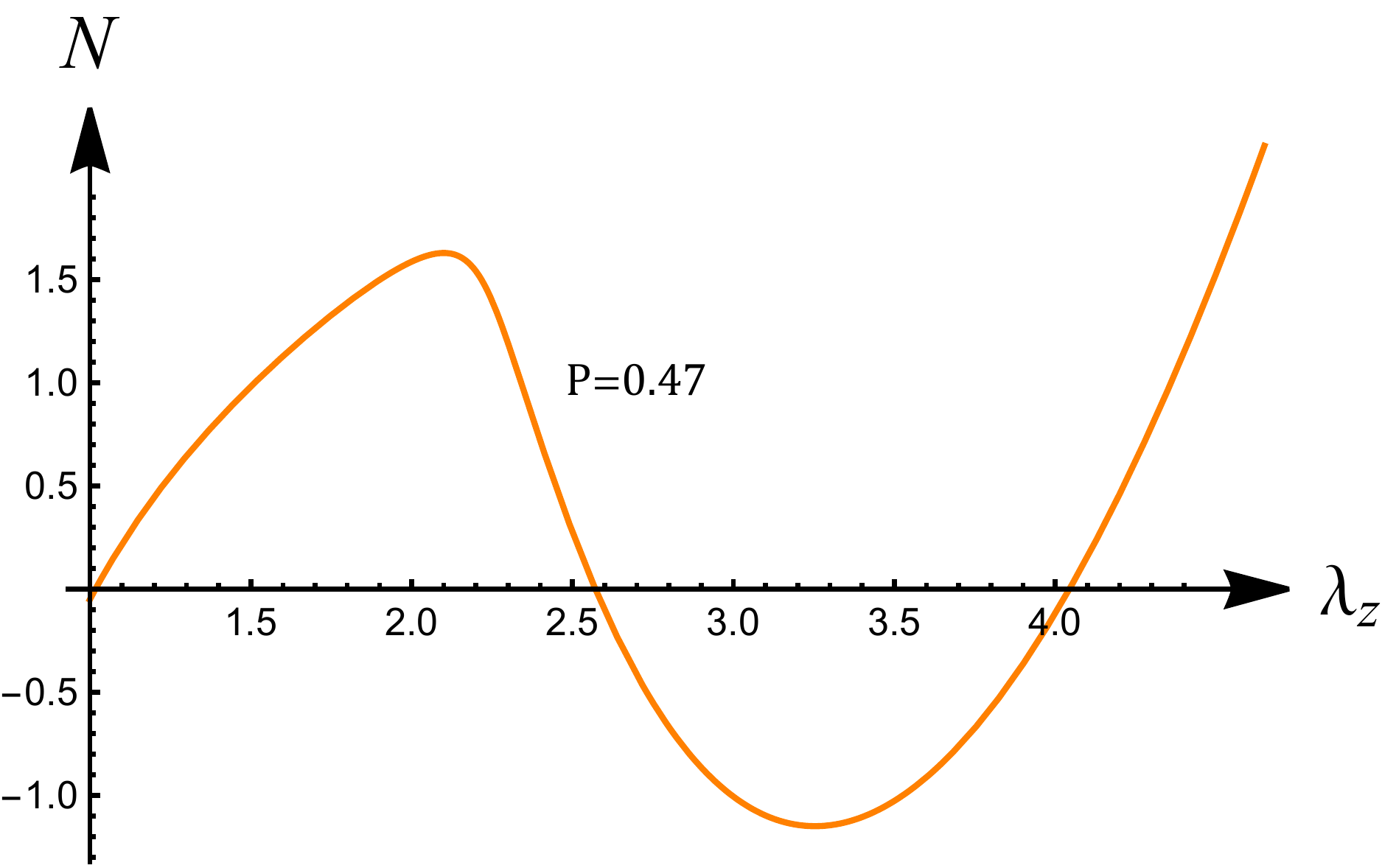}
\end{tabular}
\end{center}
\caption{$N$ versus $\lambda_{z}$ in uniform inflation corresponding to four typical values of $P$.}
\label{fig3}
\end{figure*}

\begin{figure}[ht]
\begin{center}
\begin{tabular}{ccc}
 \includegraphics[width=.4\textwidth]{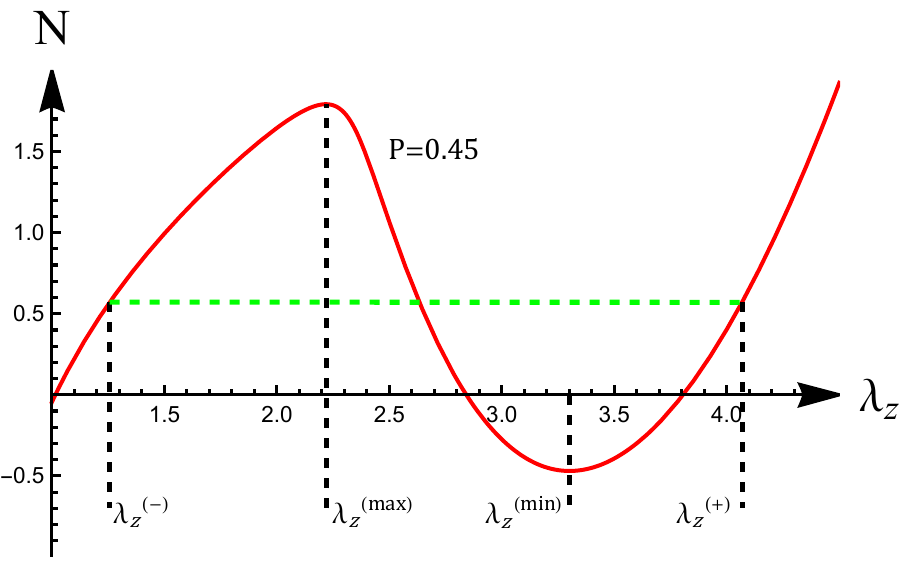}   & &\includegraphics[width=.4\textwidth]{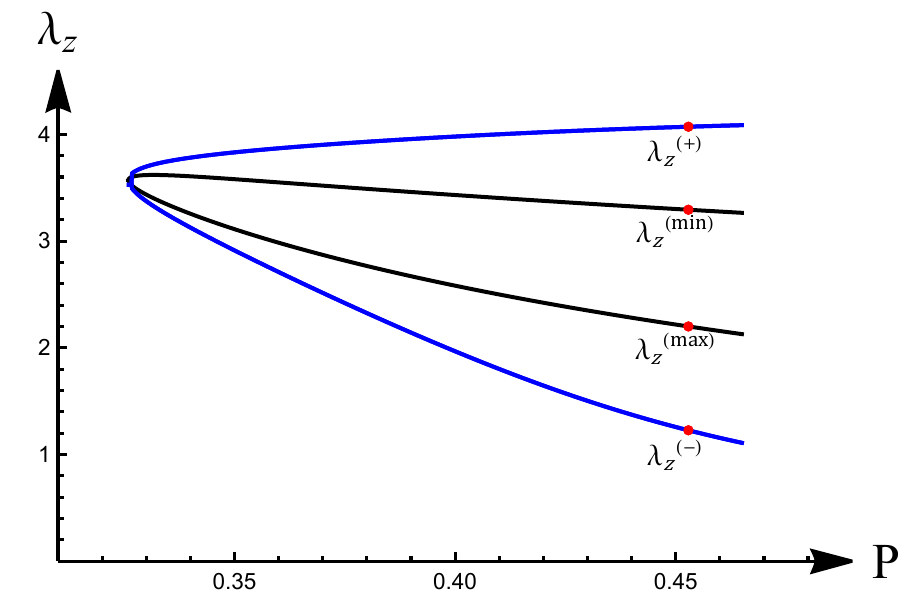}\\
 (a) & &(b)
\end{tabular}
\end{center}
\caption{(a) Definitions of $\lambda_{z}^{({\rm max})}$, $\lambda_{z}^{({\rm min})}$, $\lambda_{z}^{\rm (-)}$, $\lambda_{z}^{\rm (+)}$, and (b) their dependence on $P$.
}
\label{fig10ab}
\end{figure}

To appreciate why localised bulging becomes impossible when $P$ is less than $0.3259$ and why Abaqus simulations usually abort when $P$ reaches a large enough value, we now investigate the dependence of $N$ on $\lambda_{z}$ as $P$ is increased gradually. A typical set of results is presented in Fig.~\ref{fig3}. It is seen that when $P$ is sufficiently small, the $N$ versus $\lambda_{z}$ curve is monotonically increasing. It is only when $P$ becomes larger than the above-mentioned minimum $0.3259$ that the variation of $N$
has an ${\cal N}$ shape with a maximum followed by a minimum. This is consistent with the result that when inflation is carried out with fixed $P$ and variable $N$, localised bulging takes place when $dN/d\lambda_z=0$; see Eqn (21) in \citet{fi2015}.
For instance, when $P=0.33$, the maximum of $N$ occurs at $\lambda_z=3.4357$ (with the corresponding $\lambda_a$ equal to $1.8179$). This means that (i) when the tube is stretched with pressure fixed at $P=0.33$, localised bulging will occur when $\lambda_z$ reaches $3.4357$, and also (ii) when the tube is inflated with a pre-stretch $\lambda_z=3.4357$, localised bulging will occur when $P=0.33$.
When the variation of $N$ has an ${\cal N}$ shape, equal area rule may be applied and a Maxwell state may be obtained. Fig.~\ref{fig10ab}(b) shows the variation of the two values of $\lambda_{z}$ corresponding to the Maxwell state, together with the two values of $\lambda_{z}$ corresponding to the maximum and minimum of $N$. The notations $\lambda_{z}^{\rm max}$, $\lambda_{z}^{\rm min}$, $\lambda_{z}^{\rm (+)}$, $\lambda_{z}^{\rm (-)}$ are defined in Fig.~\ref{fig10ab}(a) where we show the variation of $N$ against $\lambda_z$ for a typical $P=0.45$. For each $P>0.3259$, Fig.~\ref{fig10ab}(b) gives the two Maxwell values $\lambda_{z}^{\rm (+)}$ and $\lambda_{z}^{\rm (-)}$. The corresponding values of $\lambda_a$, say $\lambda_{a}^{\rm (+)}$, $\lambda_{a}^{\rm (-)}$, can then be determined by solving \rr{1.1}$_1$. The larger of these two values is
the $r(0)$ appearing in Fig.~\ref{fig6}, and the collection of $(r(0), P)$ computed in this way forms the asymptote in Fig.~\ref{fig6}.

It is seen in Fig.~\ref{fig3}(d) that when $P$ becomes as large as $0.47$, the axial force $N$ corresponding to the Maxwell state is negative. When $N$ is negative or sufficiently negative depending on the wall thickness, Euler buckling or axisymmetric buckling would occur, and Abaqus simulations would eventually abort. Thus, the asymptote in Fig.~\ref{fig6} starts at the minimum $P=0.3259$ and terminates at a value approximately equal to $0.47$ at which $N=0$.

It is also seen in Fig.~\ref{fig6}(b) that the asymptote, where it exists, coincides with the solid curve corresponding to the special stretch $\lambda_z=3.5694$. This is the stretch at which $P$ attains its minimum, as mentioned earlier. This special stretch is discussed in the next section.

\begin{figure}[ht]
\begin{center}
\begin{tabular}{ccc}
 \includegraphics[width=.4\textwidth]{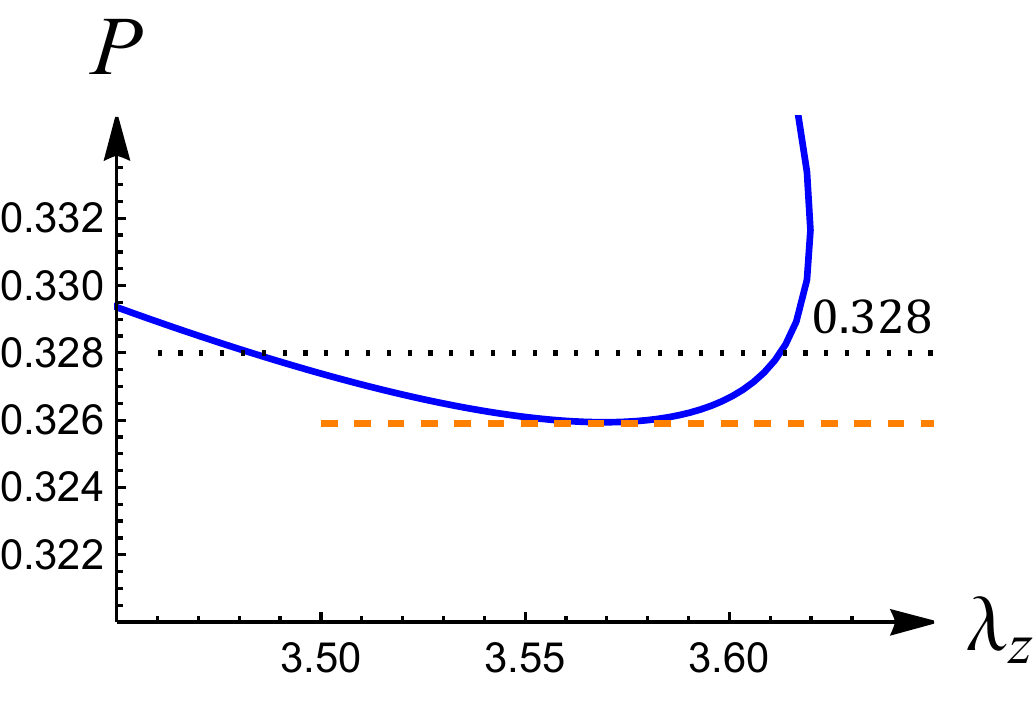} &  &\includegraphics[width=.4\textwidth]{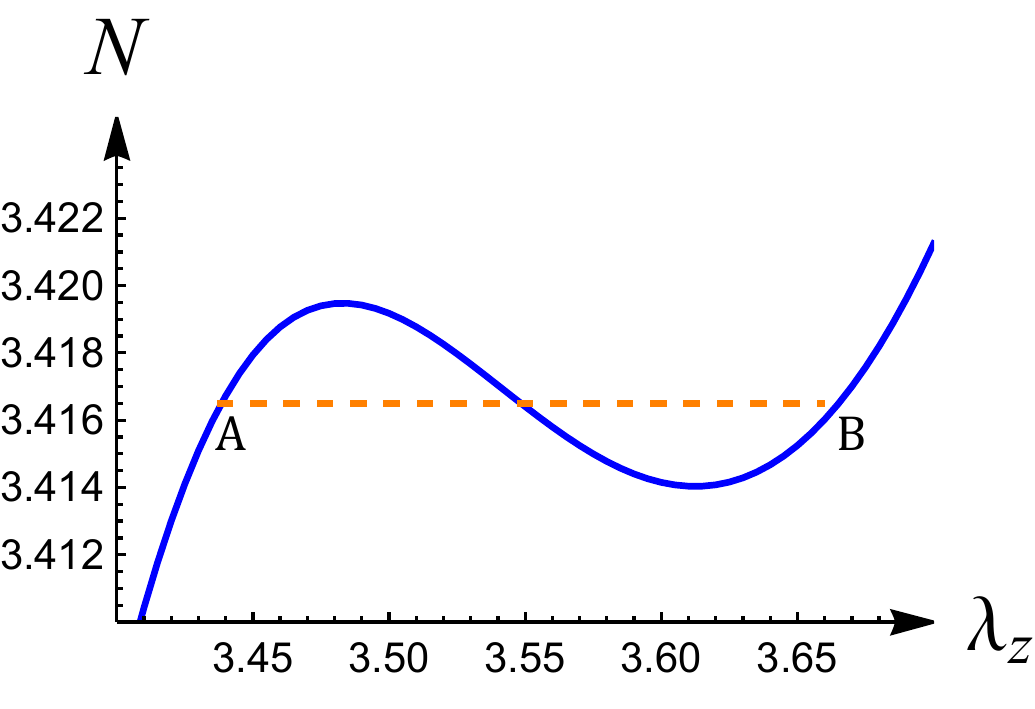}\\
 (a) & &(b)
\end{tabular}
\end{center}
\caption{(a) Close-up of the bifurcation curve near the pressure minimum in Fig.~\ref{fig7ab}(a). The dotted line corresponds to a typical value of $P$ slightly above the minimum and its intersections with the bifurcation curve (solid line) correspond to the maximum and minimum in (b).
 (b) $N$ versus $\lambda_{z}$ diagram when $P$ is equal to the value marked in (a) (slightly above its minimum). The dashed line is the Maxwell line with intersections $A$ and $B$ corresponding to a \lq\lq two phase" state.}
\label{fig11}
\end{figure}

\section{Solution near the pressure minimum}

Fig.~\ref{fig11} shows a close-up of the bifurcation curve in Fig.~\ref{fig7ab}(a) near the pressure minimum $P=0.3259$ where $\lambda_z=3.5694$ and $\lambda_a=2.0628$.
We consider the scenario whereby the tube is first subjected to an axial pre-stretch given by $\lambda_z=3.5694$,  and then inflated by increasing $P$ gradually. It is anticipated that when $P$ reaches the critical value $0.3259$, a bifurcation will take place and an inhomogeneous deformation with non-constant $\lambda_a$ and $\lambda_z$ will emerge. We denote the values of $\lambda_a$ and $\lambda_z$ near the end $Z=L/2$ by $r_{\infty}$ and $z_{\infty}$, respectively. For a tube that is sufficiently long, these values are effectively the same values as $Z \to \infty$ (the difference being exponentially small). In other words, as far as localised bulging  or a kink solution is concerned, the tube is effectively infinite provided it is sufficiently long.

 The inflation pressure corresponding to the Gent model is given by
\be P=\frac{486-486 r_\infty^4 z_\infty^2}{5 r_\infty^6 z_\infty^3+5 r_\infty^4 z_\infty^5-501
   r_\infty^4 z_\infty^3+5 r_\infty^2 z_\infty}. \la{p1} \en
This equation can be solved explicitly to express $r_\infty$ in terms of $P$ and $z_\infty$, and the solution is denoted by $r_\infty=r_\infty(P, z_\infty)$ in the subsequent analysis.

With $r_{\infty}$ and $z_{\infty}$ specified, we look for an inhomogeneous bifurcated solution that approaches this uniform state as the end $Z=L/2$ is approached.
Extending the analysis of  \citet{fpl2008}, we find that if $\lambda_a=r_{\infty}+y(Z)$, then the incremental solution $y(Z)$
satisfies the ordinary differential equation
\begin{equation}
y^{\prime 2} =\tilde{\omega}(r_\infty, z_\infty) \,y^2+ \tilde{\gamma}(r_\infty, z_\infty)\, y^3+\tilde{\kappa}(r_\infty, z_\infty)\, y^4+
O(y^5), \label{yprime1}
\end{equation}
where the three coefficients have long but explicit expressions which are not written out here for the sake of brevity. The bifurcation condition corresponds to $\tilde{\omega}(r_\infty, z_\infty)=0$ \citep{fpl2008} and the pressure minimum in Fig.~\ref{fig11} corresponds to $\tilde{\gamma}(r_\infty, z_\infty)=0$. With $r_\infty$ expressed in terms of $P$ and $z_\infty$ with the use of \rr{p1}, we rewrite equation \rr{yprime1} symbolically in the form
\begin{equation}
y^{\prime 2} ={\omega}(P, z_\infty) \,y^2+ {\gamma}(P, z_\infty)\, y^3+{\kappa}(P, z_\infty)\, y^4+
O(y^5), \label{yprime2}
\end{equation}
where, for instance,  ${\omega}(P, z_\infty)=\tilde{\omega}(r_\infty(P, z_\infty), z_\infty)$. With $z_{\infty}$ fixed, the pressure $P$ is taken to be the bifurcation parameter.

Denote the coordinates of the pressure minimum in Fig.~\ref{fig11}(a) by $(z_{\rm cr}, P_{\rm cr})$. When $z_\infty \ne z_{\rm cr}$ so that ${\gamma} \ne 0$, an explicit expression for the localised solution may be obtained from \rr{yprime2} by neglecting the quartic and higher order terms on the right hand side. It can then be deduced that the localised solution is of the bulging type if $z_\infty < z_{\rm cr}$ (corresponding to ${\gamma}(P, z_\infty)<0$), and is of the necking type if $z_\infty > z_{\rm cr}$. This switching from the bulging type to the necking type was also observed by  \citet{ylf2020} through a weakly nonlinear analysis for a tube of arbitrary thickness.  Here we focus on the special case with $z_\infty = z_{\rm cr}$.

We note, however, that despite the assumption that uniform inflation is now carried out with $z_\infty = z_{\rm cr}$, as soon as $P$ is increased beyond its minimum, a bifurcation from the homogeneous state into an  inhomogeneous state takes place and the $z_\infty$ will change in order to satisfy the condition that the total length is fixed at $l=z_{\rm cr} L$. Thus, we set
\be
 P=P_{\rm cr}+\ep P_1, \;\;\;\;z_\infty = z_{\rm cr}+ \ep^{\frac{1}{2}} z_1, \la{p2} \en
where $\ep$ is a small positive parameter, and $P_1$ and $z_1$ are constants. On substituting these expansions into the solution $r_\infty=r_\infty(P, z_\infty)$ of  \rr{p1}, we obtain
\be
r_\infty=r_{\rm cr}+ \ep^{\frac{1}{2}} r_1, \la{rrr} \en
where $r_{\rm cr}=2.0628, \; r_1=3.3209 z_1$.

Expanding the right hand side of \rr{yprime2} in terms of $\ep$, we obtain
\begin{equation}
y^{\prime 2} =\ep ( {\omega}_1 P_1+\frac{1}{2} {\omega}_{22} z_1^2  ) \,y^2+ \ep^{\frac{1}{2}} z_1 {\gamma}_2 \, y^3+{\kappa}_0\, y^4+
O(\ep^2 y^2, \ep y^3, y^5), \label{yprime3}\en
where ${\omega}_1=\paa {\omega}/\paa P$, ${\omega}_{22}=\paa^2 {\omega}/\paa z_\infty^2$,  ${\gamma}_2=\paa {\gamma}/\paa z_\infty$, ${\kappa}_0={\kappa}(P_{\rm cr}, z_{\rm cr})$, all partial derivatives being evaluated at the pressure minimum. In obtaining \rr{yprime3} we have made use of the property that $\paa {\omega}/\paa z_\infty=0$ at the pressure minimum. This property also explains the scaling in \rr{p2}$_2$.

It can be seen that the first four terms on the right hand side of \rr{yprime3} are all of the same order as the left hand side if $y(Z)=O(\ep^{\frac{1}{2}})$ and $y(Z)$ varies on a long spatial scale of order $\ep^{-\frac{1}{2}}$. In terms of the scaled variables
\be Y(s)=y(Z)/\ep^{\frac{1}{2}}, \;\;\;\; s=\ep^{\frac{1}{2}} Z, \la{scaled} \en
the amplitude equation \rr{yprime3} reduces to
\be
Y^{\prime 2} =  ( {\omega}_1 P_1+\frac{1}{2} {\omega}_{22} z_1^2  )  \,Y^2+z_1 {\gamma}_2 \, Y^3+{\kappa}_0\, Y^4
 = {\kappa}_0\, Y^2 (Y- a)(Y-b), \label{yprime4}\en
after we have neglected terms of order $\ep^{5/2}$ or higher, where $a$ and $b$ are the two roots of
$$
{\kappa}_0\, Y^2+ z_1 {\gamma}_2 \, Y+ ( {\omega}_1 P_1+\frac{1}{2} {\omega}_{22} z_1^2  ) =0. $$
We consider the special case when these two roots are repeated, which occurs when
\be z_1=\pm \sqrt{\frac{4{\kappa}_0 {\omega}_1 P_1 }{ {\gamma}_2^2-2 {\kappa}_0 {\omega}_{22} }}, \la{jan1} \en
and the roots are
\be a=b=-\frac{z_1 {\gamma}_2}{2{\kappa}_0 }. \la{jan2} \en
In this case, equation \rr{yprime4} reduces to $Y'=\sqrt{{\kappa}_0} Y (Y-a)$, where we have assumed, without loss of generality, that $Y'<0$ so that the solution {\it decays} to the constant state as the end $Z=L/2$ is approached. This differential equation can be integrated to yield the explicit kink-wave solution \citep{giudici2020, fjg2021}
\be Y(s)=\frac{a}{1+ \, {\rm exp} ( a \sqrt{{\kappa}_0} s)}= \frac{a}{2} -\frac{a}{2} \tanh (\frac{1}{2} a \sqrt{{\kappa}_0} s), \la{max} \en
where the constant of integration has been chosen such that $Y(0)=a/2$.

For the Gent material model under consideration, we find that
\be \omega_1=-2.3989, \omega_{22}=2.3961, \gamma_2=0.2405, \kappa_0=0.0181. \la{num0} \en
To conform with the assumption $Y'<0$, we require $a>0$ so that we take the minus choice in \rr{jan1} for $z_1$. As a result, as $Z \to \pm \infty$, we have $Y(s) \to 0$ and $a$, respectively. Thus, the non-trivial solution represents a \lq\lq two-phase" deformation (or a kink-wave solution) with the \lq\lq two-phases" given by
$$
r_\infty= r_{\rm cr}+ \ep^{\frac{1}{2}} r_1+\ep^{\frac{1}{2}} Y(\infty), \;\;\;\; {\rm or}\;\;\;\; r_{\rm cr}+ \ep^{\frac{1}{2}} r_1+\ep^{\frac{1}{2}} Y(-\infty), $$
or more precisely,
\be
r_\infty= r_{\rm cr}+ \ep^{\frac{1}{2}} r_1, \;\;\;\; {\rm or}\;\;\;\; r_{\rm cr}+ \ep^{\frac{1}{2}} r_1+\ep^{\frac{1}{2}} \frac{  {\gamma}_2}{{\kappa}_0 } \sqrt{\frac{{\kappa}_0 {\omega}_1 P_1 }{ {\gamma}_2^2-2 {\kappa}_0 {\omega}_{22} }}. \la{jan5} \en
The corresponding values of $z_\infty$ are
\be
z_\infty= z_{\rm cr}-   \sqrt{\frac{4{\kappa}_0 {\omega}_1 (P-P_{\rm cr}) }{ {\gamma}_2^2-2 {\kappa}_0 {\omega}_{22} }}, \;\;\;\; {\rm or}\;\;\;\; z_{\rm cr}+ (-2+ \frac{  {\gamma}_2}{{\kappa}_0 } \cdot \frac{z_1}{r_1} ) \sqrt{\frac{{\kappa}_0 {\omega}_1 (P-P_{\rm cr}) }{ {\gamma}_2^2-2 {\kappa}_0 {\omega}_{22} }}, \la{jan6} \en
where the ratio $z_1/r_1$ is given just below \rr{rrr}.

As a consistency check, we note that the two values of $z_\infty$ given by \rr{jan6} should correspond to the coordinates of points $A$ and $B$ in Fig.~\ref{fig11}(b) where the dashed line cuts the curve into equal areas. For values of $P$ and $z_\infty$ close to $P_{\rm cr}$ and $z_{\rm cr}$, respectively, the $N$ has the Taylor expansion
$$
N=N(P_{\rm cr}, z_{\rm cr})+ N_p (P-P_{\rm cr})+\frac{1}{2} N_{pp} (P-P_{\rm cr})^2 $$
\be \hspace{1.5cm} + N_{pz} (P-P_{\rm cr}) (z_\infty-z_{\rm cr})+\frac{1}{6} N_{zzz} (z-z_{\rm cr})^3+\cdots, \la{jan7} \en
where
$$ N_p=\frac{\paa N}{\paa P}, \;\;\;\;  N_{pp}=\frac{\paa^2 N}{\paa P^2}, \;\;\;\;
N_{pz}= \frac{\paa^2 N}{\paa P \paa z_\infty}, \;\;\;\; N_{zzz}=\frac{\paa^3 N}{\paa z_\infty^3},$$
with all the partial derivatives evaluated at the pressure minimum. It then follows that to leading order the dashed line in  Fig.~\ref{fig11}(b) is given by
\be
N=N_{m} \equiv N(P_{\rm cr}, z_{\rm cr})+ N_p (P-P_{\rm cr})+\frac{1}{2} N_{pp} (P-P_{\rm cr})^2, \la{jan7a} \en
and the  points $A$ and $B$ have coordinates
\be
z_\infty^{{\rm B}, {\rm A}}=z_{\rm cr}\pm \sqrt{- 6 (P-P_{\rm cr})N_{pz} /N_{zzz}}. \la{jan9} \en
Consistency between this result and \rr{jan6} requires that
\be
 - \frac{3  N_{pz}}{N_{zzz}} =\frac{2{\kappa}_0 {\omega}_1 }{ {\gamma}_2^2-2 {\kappa}_0 {\omega}_{22} }, \;\;\;\;\;\;
\frac{  {\gamma}_2}{{\kappa}_0 } \cdot \frac{z_1}{r_1}=4. \la{jan10} \en
We have verified numerically that these relations are indeed satisfied.


Combining the simulation results in Fig.\,\ref{fig6}(b), the theoretical results in Fig.\,\ref{fig10ab}(b), and the analysis in this section so far, we may now summarise the evolution of the deformation in an inflated tube with fixed ends as follows.  If inflation is carried out with a pre-stretch $\lambda_z<z_{\rm cr}$, then increasing $P$ with the fixed $\lambda_z$ follows a horizontal line in Fig.\,\ref{fig10ab}(b). When this line intersects the curve marked \lq\lq $\lambda_z^{\rm (max)}$", a localised bulge will initiate. If inflation is carried out with mass control, the pressure will decrease to a minimum and then rise again, as shown in Fig.\,\ref{fig6}(b). However, if it is under pressure control (so that pressure is not allowed to decrease), a snap-through will take place and the deformation will jump to a \lq\lq two-phase" deformation with the values of $\lambda_z$ in the two \lq\lq phases" given by $\lambda_z^{(+)}$ and $\lambda_z^{(-)}$, respectively; the latter two values can be read off from Fig.\,\ref{fig10ab}(b) by drawing a vertical line at the pressure assigned. If the pressure is increased further after the snap-through, the values of $\lambda_z^{(+)}$ and $\lambda_z^{(-)}$ in the two \lq\lq phases" will change continuously by staying on the same branches in Fig.\,\ref{fig10ab}(b). On the other hand, if inflation is carried out with $z_{\rm cr}<\lambda_z<z_{z\rm max}$, the behaviour is similar except that the localisation is of the necking type and the intersection is with the curve marked \lq\lq $\lambda_z^{\rm (min)}$" in Fig.\,\ref{fig10ab}(b). In the exceptional case when inflation is carried out with $\lambda_z=z_{\rm cr}$ exactly, the bifurcation is exceptionally supercritical: when $P$ reaches the minimum in Fig.\,\ref{fig10ab}(b), the deformation {\it smoothly} evolves into a \lq\lq two-phase" deformation without going through a snap-through, as can be seen in Fig.\,\ref{fig6}(b). This two \lq\lq phase" deformation is described analytically by \rr{max} when $P$ is only slightly larger than its minimum. Further away from this minimum, the values of $\lambda_z^{(+)}$ and $\lambda_z^{(-)}$ in the two \lq\lq phases" are given by the same branches in Fig.\,\ref{fig10ab}(b) although an analytical expression for the narrow transition region no longer exists. The proportion of each \lq\lq phase" is determined such that the total length of the tube remains at $l=\lambda_z L$.


%
%
%

\section{Conclusion}
The current study was motivated by two considerations. Firstly, it is observed that the Ogden material model and the Gent or Gent-Gent material model make similar and realistic predictions for localised bulging when the axial pre-strain (pre-stretch minus unity) or the fixed axial force is sufficiently small, but their predictions diverge for large values of axial pre-strain or   tension. It is then of interest to investigate how an actual rubber tube would behave under large pre-strains. Secondly, the entire inflation process for the case of fixed axial force is well understood both theoretically and experimentally, but this does not seem to be the case when the ends of the tube are fixed during inflation. Previous experiments have shown that after the initiation of a localised bulge the pressure would reach a minimum and then rise again, but this ascending branch has not previously been described analytically.

We carried out experiments using a typical commercially available rubber tube and confirmed that there does indeed exist a maximum pre-stretch above which localised bulging will not occur under inflation.  The Gent material model can describe this property both qualitatively and quantitatively. This is probably because the existence of the maximum pre-stretch for localised bulging is closely related to the finite chain extensibility that the Gent model was designed to model. The Gent-Gent model can predict this property qualitatively. It is also possible to predict this property quantitatively by re-calibrating the Gent-Gent model and requiring $C_2$ to be sufficiently small in fitting the model to experimental data, guided by the results in Fig.\,\ref{newfig2}(b), but we did not
pursue this possibility further. We cannot, however, conclude that this property is shared by all other rubber or rubber-like materials.
Thus, the current study only serves to show that finding a tractable material model that can serve all purposes is extremely challenging, if not impossible.

Our analysis of the above-mentioned ascending branch has been inspired by the recent study on the localised bulging/necking of a solid cylinder under surface tension \citep{fjg2021}, which in turn has benefitted from previous studies on the problem of localised bulging in an inflated rubber tube without surface tension. In view of what has been achieved in the latter paper, we anticipated that the ascending branch should tend to an asymptote that is independent of the tube length or the pre-stretch. This was verified by both analysis and Abaqus simulations in the current study.

It is previously known that localised bulging is also possible when the tube is extended/stretched at a fixed pressure, and it occurs when the tension $N$ reaches the maximum corresponding to uniform extension.
By presenting the bifurcation condition in the $(\lambda_z, P)$-plane, it is revealed that there exists a minimum pressure below which localized bulging would be impossible if the tube were to be continuously stretched axially. This minimum pressure is attained at $\lambda_z=z_{\rm cr}=3.5694$ when the Gent material model is used. The latter stretch value is very close to the stretch maximum  $3.6199$ above which localised bulging becomes impossible when the tube is inflated by an internal pressure. For inflation with fixed ends and a pre-stretched $\lambda_z<3.6199$, the first bifurcation corresponds to localized bulging if $\lambda_z<z_{\rm cr}$ and localised necking if $\lambda_z>z_{\rm cr}$, and the bifurcation is sub-critical in both cases. In the exceptional case when inflation is carried out with $\lambda_z=z_{\rm cr}$, the tube will deform smoothly into a \lq\lq two-phase" state when the pressure is increased across the critical value predicted by the bifurcation condition, and the bifurcation is exceptionally super-critical.

\subsection*{Acknowledgements}
This work was supported by the National Natural Science Foundation of China (Grant Nos 12072224, 12002067). The Abaqus simulations were carried out on TianHe-1 (A) at the National Supercomputer Center in Tianjin, China.

\end{document}